\documentclass[reprint,superscriptaddress,floatfix,amsmath,amssymb,pra,aps,showpacs,twocolumn]{revtex4-2}
\usepackage{amsmath,amsfonts,amssymb,amsthm,graphics,graphicx,epsfig,bbm}
\usepackage[colorlinks=true,citecolor=blue,linkcolor=blue,urlcolor=blue]{hyperref}
\usepackage[usenames]{color}
\usepackage{graphicx}
\usepackage[export]{adjustbox}
\usepackage{subfigure}
\usepackage{amsmath}
\usepackage{epsfig}
\usepackage{dcolumn}
\usepackage{bm}
\usepackage{color}
\usepackage{times}
\usepackage{epstopdf}
\usepackage{amssymb}
\usepackage{amstext}
\usepackage{latexsym}
\usepackage{hyperref}
\usepackage{amsfonts}
\usepackage{psfrag}
\usepackage[utf8]{inputenc}
\usepackage{graphicx}
\usepackage{amsmath}
\usepackage{amssymb}
\usepackage{dsfont}
\usepackage{tensor}
\usepackage{color}
\usepackage{ulem}
\usepackage{upgreek}
\usepackage{braket}
\usepackage{algpseudocode}
\usepackage{algorithm}
\usepackage{orcidlink}  
\usepackage{multirow}

\UseRawInputEncoding

\begin{document}

\title{Memory effects in pulsed optomechanical systems}

\author{Hachisko Tapia-Maureira\orcidlink{0009-0003-8300-6503}}
\email{email: hachiskoo@gmail.com}
\affiliation{Centro Multidisciplinario de F\'isica, Universidad Mayor, ´
Camino la Piramide 5750, Huechuraba, Santiago, Chile}

\author{Bing He\orcidlink{0000-0002-7169-0446}}
\affiliation{Centro Multidisciplinario de F\'isica, Universidad Mayor, ´
Camino la Piramide 5750, Huechuraba, Santiago, Chile}

\author{Massimiliano Di Ventra\orcidlink{0000-0001-9416-189X}}
\email{email: diventra@physics.ucsd.edu}
\affiliation{Department of Physics, University of California San Diego, La Jolla, CA 92093}

\author{Ariel Norambuena\orcidlink{0000-0001-9496-8765}}
\email{ariel.norambuena@usm.cl} 
\affiliation{Departamento de F\'isica, Universidad T\'ecnica Federico Santa Mar\'ia, Casilla 110 V, Valpara\'iso, Chile}

\date{\today}

\begin{abstract}
Memory, understood as time non-locality, is a fundamental property of any physical system, whether classical or quantum, and has important applications in a wide variety of technologies. In the context of quantum technologies, systems with memory can be used in quantum information, communication, and sensing. Here, we demonstrate that cavity optomechanical systems driven by 
a pulsed laser can operate as programmable quantum memory elements. By engineering the adiabatic and non-adiabatic pulses, 
particularly the Gaussian and sinusoidal drives, we induce and control diverse memory phenomena such as dynamical hysteresis, quantized phononic transitions, and distinct energy-storing responses. Within a mean-field approach, we derive the analytical and numerical criteria under which the photonic and phononic observables manifest the memory effects in strongly driven regimes. The memory effects are quantified through a dimensionless geometric form factor, which provides a versatile metric to characterize the memory efficiency. Our protocol is readily compatible with the current optomechanical platforms, highlighting the new possibilities for advanced memory functionalities in quantum technologies.
\end{abstract}

\maketitle

\section{Introduction}

Although the word ``memory'' may acquire different meanings in diverse contexts, in the present work we will refer to the non-equilibrium property of a physical system such that when its state is perturbed by an external signal, the effect of such a perturbation persists in time, namely it will affect the state of the system at a later time~\cite{DiVentraBook2023}. Such a property, which we could also term ``time non-locality'', plays a central role across scientific disciplines, underpinning information storage and retrieval in biological, classical, and quantum systems. A classic example is magnetic hysteresis, where the system's current state explicitly depends on its past dynamics, enabling non-volatile memory functionalities crucial for the technological advances in data storage~\cite{Hoffmann2008,Pershin2011}.

Recently, volatile and non-volatile memory elements~\cite{DiVentra2009,Martinez-Rincon2011,Bagheri2011} have become the key building blocks in neuromorphic and reservoir computing, where information processing relies on transient dynamics and nonlinear responses rather than static configurations~\cite{Marković2020,Lukosevicius2009}. This principle is embodied by so-called ``memdevices'', including memristors, memcapacitors, and meminductors~\cite{DiVentra2009}, which showcase the history-dependent responses under time-dependent driving. Also, mechanical systems with time non-local responses, such as mem-dashpot~\cite{memdashpot}, mem-inerter~\cite{meminerter}, and mem-spring~\cite{memspring} exhibit memory in their behavior. Such devices have opened new avenues for the next-generation computing architectures, particularly in the platforms featuring intrinsic nonlinearities and complex dynamical responses~\cite{MemComputingbook,Pershin2011}.

In open quantum systems, memory is typically associated with environment-induced non-Markovian dynamics, characterized by completely-positive (CP) divisibility breakdown in quantum dynamical maps~\cite{Wolf2008,BLP2009}. A process \( \Lambda \) is CP-divisible if it satisfies \( \Lambda(t,0) = \Lambda(t,s)\Lambda(s,0) \) for all \( s \leq t \), with each intermediate map being completely positive and trace-preserving~\cite{Fanchini2021}. Violating this property indicates non-Markovian behavior~\cite{deVega2017}, which can be quantified through the measures based on trace distance~\cite{BLP2009,Rivas2014} or entropic correlations~\cite{Rivas2010,Fanchini2014}. Recent studies have clarified the connections and hierarchy between these approaches~\cite{Li2023}.

Beyond environment-induced memory effects, memory can also emerge from the intrinsic dynamics of driven quantum systems. This aligns with the classical notion of hysteresis, emphasizing the role of dynamical responses to external stimuli. In particular, dynamical hysteresis in quantum observables under pulsed or periodic control can reveal underlying memory effects~\cite{QMEM_cQED, Qmem_entagled, Qmem_Polariton}. In this context, quantum memristive behavior has now been theoretically predicted or experimentally demonstrated across various platforms, including superconducting circuits~\cite{Peotta,QMEM_cQED,Qmem_entagled}, light-matter systems~\cite{Qmem_Polariton}, quantum dots~\cite{Qmem_dots}, trapped ions~\cite{QMem_TrappedIon}, bosonic reservoirs~\cite{Emem_tunable}, quantum processors~\cite{Qmem_Qcomp}, and beam splitter in photonic devices~\cite{PQM_1,PQM_1experimental,PQM_2experimental,PQM_3}.

Among these platforms, optomechanical (OM) systems are highly versatile candidates for engineering the memory effects at quantum and mesoscopic scales. The interplay between the radiation-pressure coupling and laser-driven control enables precise manipulation of photonic and phononic modes~\cite{Aspelmeyer2014}, fostering the exploration of the memory phenomena in light-mechanics systems. So far, OM systems have supported the applications in sensing~\cite{OM_BHe1}, metrology, and quantum information processing~\cite{OM_BHe2,OM_BHe3,OM_BHe4,OM_BHe6_Ent,OM_Bhe8_BS}, and display rich dynamical behaviors including cooling~\cite{OM_BHe4}, amplitude locking~\cite{OM_BHe1,OM_BheMO_frozen}, and chaotic behavior~\cite{GuiLei2022}. Although memory in OM systems has been explored through feedback mechanisms~\cite{Vitali2018} and structured reservoirs~\cite{Groblacher2015}, the emergence of memory via coherent, pulsed driving remained 
to be largely unaddressed.

\begin{figure}[ht!]
\centering
\includegraphics[width = 1 \linewidth]{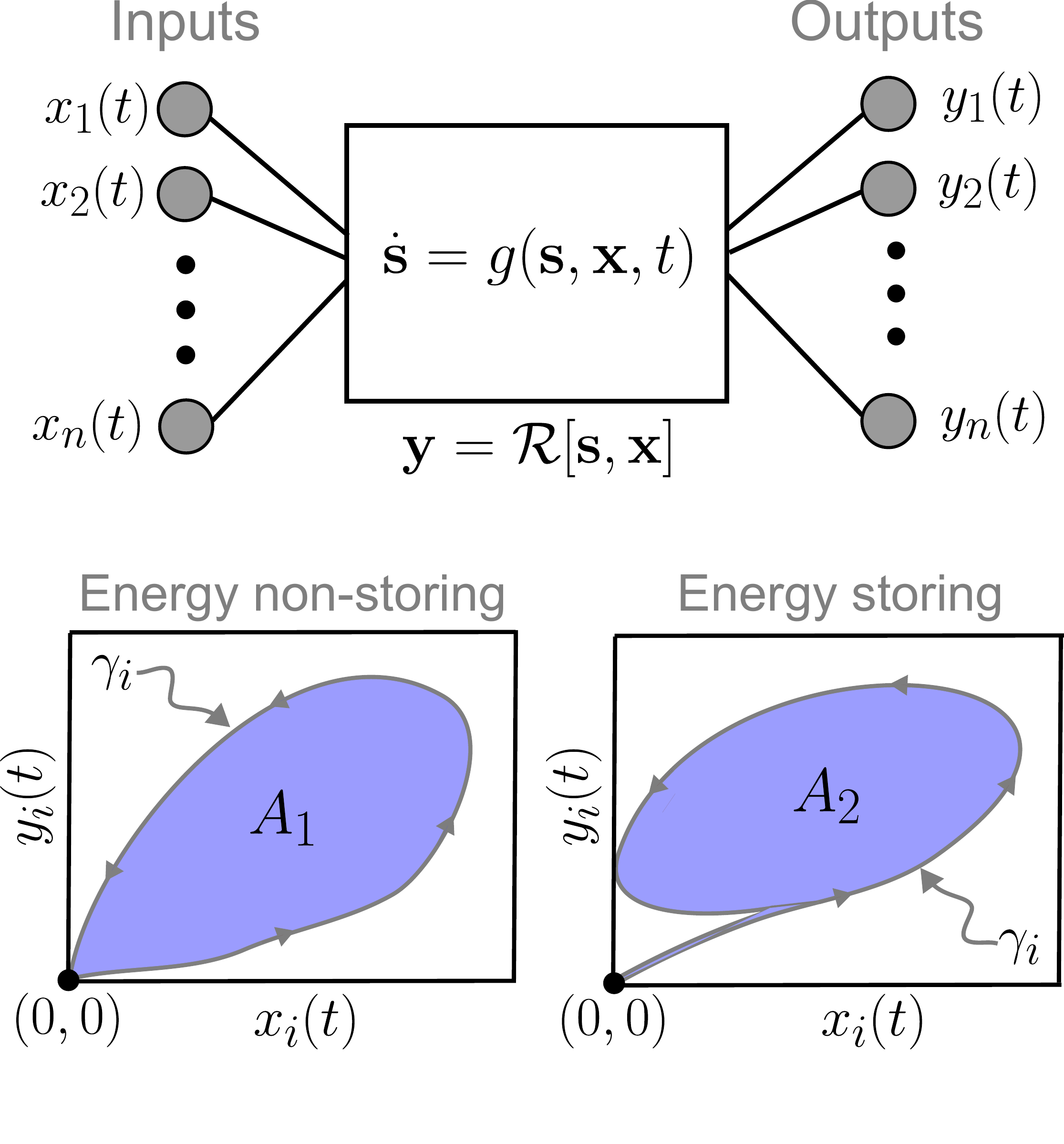}
\caption{Scheme of the input-output framework to compute memory responses in dynamical systems ruled by the equation $\dot{\mathbf{s}} = g(\mathbf{s},\mathbf{x},t)$, where $\mathbf{x}(t) = (x_1(t),x_2(t),...,x_n(t))$ are the inputs. The set of outputs are denoted by the vector $\mathbf{y}(t) = (y_1(t),y_2(t),...,y_n(t))$. Memory effects can be understood in the $x_i-y_i$ plane when a periodic driving $x_i(t+T)=x_i(t)$ is applied. Energy-storing and energy non-storing responses are illustrated, where the area $A_i \neq 0$ ($i=1,2$) is a signal of memory. Here, the origin $(0,0)$ illustrates the main difference between both responses.}
\label{fig:Figure1}
\end{figure}

In this work, we propose a novel class of OM memory protocols based on tailored periodic pulsed laser driving. For these diverse and tunable memory effects, we use a geometric criterion based on the form factor to quantify the memory using input-output trajectories~\cite{Hernani2024}. By means of a mean-field theoretical framework validated through numerical simulations, we demonstrate how adiabatic and non-adiabatic pulses yield distinctive memory behaviors, specifically under the Gaussian and sinusoidal drives, with the possibility of transitioning from ``energy storing'' to ``energy non-storing'' responses. Furthermore, we benchmark these memory effects with other proposals, identifying the optimal conditions for OM memory encoding. Our findings position cavity optomechanics as a versatile and experimentally viable platform for programmable quantum memory technologies, bridging fundamental quantum dynamics and practical applications.

\section{Memory effects and energy storing} \label{MemoryEffects}

Consider a physical system \( \mathcal{S} \) described by internal state variables \( \{s_i(t)\}_{i=1}^{n} \), forming a state vector \( \mathbf{s}(t) = (s_1,\dots, s_n)^T \in \mathbb{R}^n \), which can represent classical or quantum observables. An external input \( \mathbf{x}(t) \in \mathbb{R}^m \) drives the system, typically a set of control fields such as lasers, microwaves, voltages, electric fields, or mechanical forces. The internal dynamics is governed by
\begin{eqnarray} \label{Dynamics}
    \dot{\mathbf{s}} &=& g(\mathbf{s},\mathbf{x},t), \quad \mathbf{s}(t_0)= \mathbf{s}_0,
\end{eqnarray}
where \( g(\cdot) \) is the evolution function and \( \mathbf{s}_0 \) the initial state at time $t=t_0$. A first-order form is always attainable even for higher-order systems, as in Hamiltonian mechanics.

To quantify the system's response, we define an output function \( \mathbf{y}(t) \in \mathbb{R}^m \), which may correspond to classical measurements or quantum expectation values. This output can be modeled as a functional response
\begin{equation} \label{FunctionalResponse}
    \mathbf{y}(t) = \mathcal{R}\left[\{ \mathbf{s}(t') \}_{t' \leq t}, \{ \mathbf{x}(t') \}_{t' \leq t} \right],
\end{equation}
where \( \mathcal{R} \) may be an integral operator, convolution, or nonlinear mapping, highlighting that both the input and internal states influence the output over past times $t' \leq t$.

Following Ref.~\cite{DiVentraBook2023}, memory in physical systems arises from their inherent dynamical nature. A specific formulation of time non-local responses is given by Di Ventra and Pershin~\cite{DiVentraBook2023}
\begin{eqnarray} \label{Non-Local}
    y(t) &=& \int_{t_0}^{t} f(\mathbf{s},t,t')\, x(t')\, dt',
\end{eqnarray}
where \( f(\mathbf{s},t,t') \) is the response kernel. This structure captures memory via its explicit dependence on two time variables \( t \) and \( t' \), which arise naturally in quantum mechanics through Green's and two-point correlation functions in open quantum systems. A schematic representation of the input-output model is illustrated in the upper panel of Fig.~\ref{fig:Figure1}.

In the regime in which \( f(\mathbf{s},t,t') = \delta(t - t') f(\mathbf{s},x,t) \), Eq.~\eqref{Non-Local} reduces to \( y(t) = f(\mathbf{s},x,t)x(t) \), consistent with classical memristive systems, if $x(t)$ is the voltage, $y(t)$ the current, and $f(\cdot)$ the conductance~\cite{DiVentra2013,PQM_1experimental}. In such systems, plotting the trajectory \( \gamma(t) = (x(t), y(t)) \) under periodic driving can reveal memory effects when the area of $\gamma(t)$ is nonzero. Notably, this case implies that \( (x,y) = (0,0) \) is always a solution. This means that when the input is zero, the output is also zero, and no energy can be stored in the system, even temporarily. If the response persists after the input vanishes, so \( (0,0) \) is not part of the trajectory, then some energy may be (temporarily) stored in some degrees of freedom of the system. Therefore, we will use the concepts ``energy non-storing'' and ``energy storing'' to classify these memory responses, as shown in the lower panel of Fig.~\ref{fig:Figure1}. We will demonstrate in Sec.~\ref{Control-Energy-Storing} that, via external driving engineering, we can tune our proposed OM system to showcase both types of responses.

The previous observations motivate a geometric criterion for quantifying memory. Define an input-output trajectory \( \gamma_i(t) = (x_i(t), y_i(t)) \) over a period \( T \), such that $t \in \mathcal{I}=[t_0, t_0+T]$. If \( x_i(t) = x_i(t+T) \), the area enclosed by the trajectory can be computed via Green's theorem
\begin{equation}
 A = \left| \oint_{\gamma_i} x_i\, dy_i \right| = \left| \oint_{\gamma_i} y_i\, dx_i \right|.   
\end{equation}

\begin{figure}[ht!]
\centering
\includegraphics[width = 1 \linewidth]{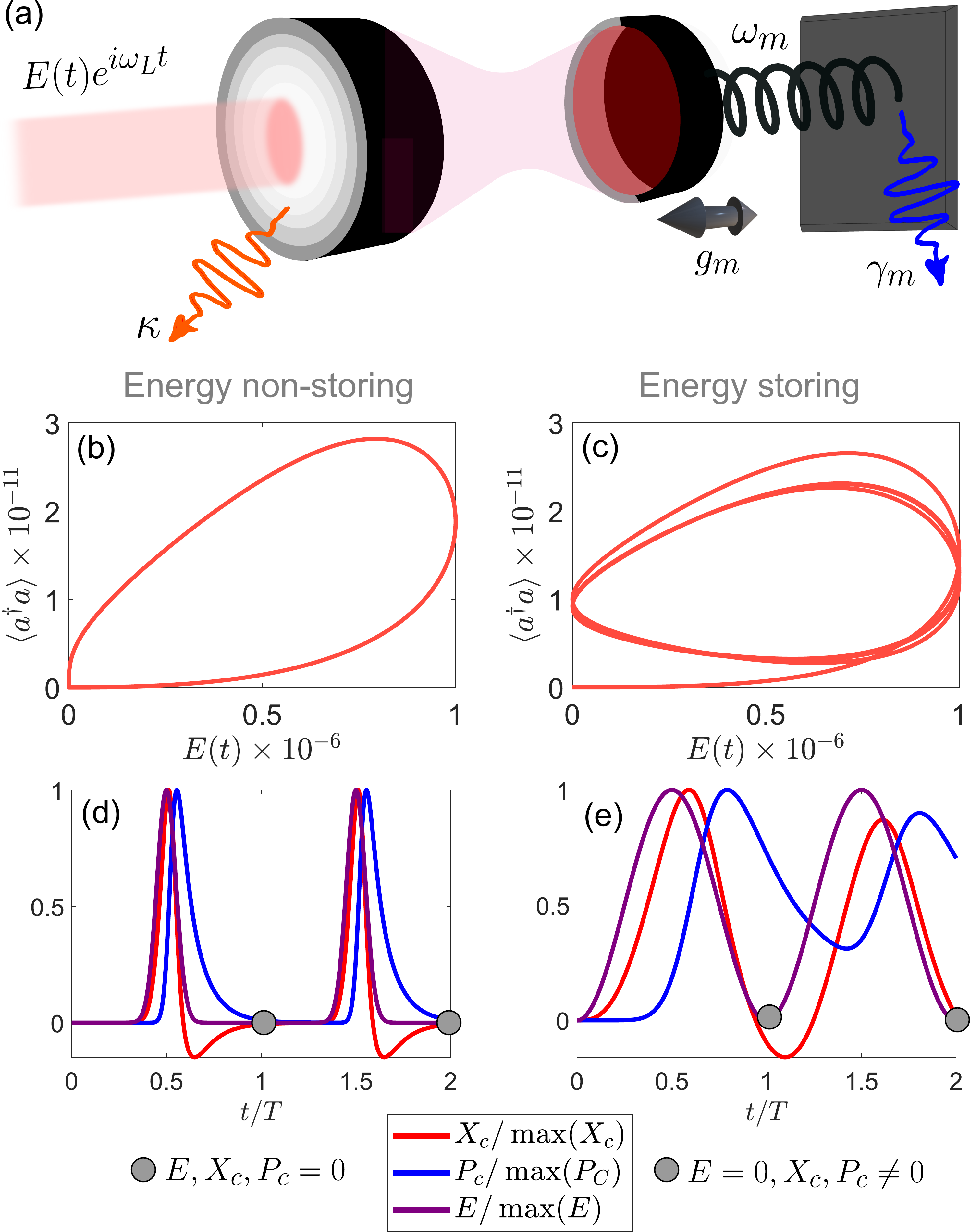}
\caption{(a) Schematic of the optomechanical setup driven by a laser field with time-dependent amplitude \( E(t) \) and frequency \( \omega_L \). The system is characterized by mechanical frequency \( \omega_m \), optomechanical coupling \( g_m \), cavity damping \( \kappa \), and mechanical damping \( \gamma_m = \omega_m/Q \). (b) and (c) show energy non-storing and storing memory effects, respectively, for different input pulses $E(t) = \sum_{n \in \rm odd} E_0 \exp[-(t - nt_s)^2 / (2\sigma^2)]$ and $E(t) = E_0 \sin^2(\omega t)$, respectively. (d) and (e) show the normalized input and photonic quadratures during two cycles for the Gaussian train and square-sinusoidal signals, respectively. Parameters: $E_0 =10^6\kappa$, $t_s = 5/\kappa$, $\sigma = t_s/10$, $\omega = 0.075\omega_m$, $\omega_m = 20 \kappa$, $Q = 10^{4}$, $g_m = 10^{-5}\kappa$.}
\label{fig:Figure2}
\end{figure}

For a memoryless system, such as \( y(t) = R x(t) \) with $R$ a constant, the area is zero. A nonzero area signals a history-dependent response. Additionally, the perimeter \( P = \oint_\gamma (\dot{x}_i^2 + \dot{y}_i^2)^{1/2} dt \) provides complementary geometric information, assuming the same physical units for $x_i$ and $y_i$. The dimensionless form factor in a cycle $t \in \mathcal{I}$ is defined as~\cite{Hernani2024}
\begin{equation} \label{FormFactor}
    \mathcal{F} = \frac{4 \pi A}{P^2} = 4\pi {\displaystyle{\left| \int_{t \in \mathcal{I}}^{} x_i \dot{y}_i\, dt \right|} \over \displaystyle{\int_{t \in \mathcal{I}} \sqrt{\dot{x}_i^2 + \dot{y}_i^2} dt} }. 
\end{equation}

The form factor $\mathcal{F}$ quantifies how efficiently a given input-output trajectory encodes memory, satisfying $0 \leq \mathcal{F} \leq 1$ due to the isoperimetric inequality. Specifically, $\mathcal{F}=0$ indicates a memoryless response, while $\mathcal{F}=1$ corresponds to an ideal circular hysteresis loop, representing maximal memory encoding efficiency. This geometric measure emphasizes the causal relationship between inputs and outputs, rather than details of the internal phase-space dynamics.

\section{Pulsed Optomechanical System} \label{pulsedOM}

We now explicitly demonstrate how memory effects naturally emerge from the intrinsic nonlinear dynamics of optomechanical (OM) systems under pulsed laser driving. We consider a dissipative OM system driven by an external laser field with a time-dependent amplitude \( E(t) \). In the rotating frame defined by the laser frequency, the system Hamiltonian is given by ($\hbar = 1$)

\begin{equation}
    H = \Delta a^{\dagger}a + \omega_m b^{\dagger}b - g_m a^{\dagger}a(b^{\dagger}+b) + i E(t)(a^{\dagger} - a),
\end{equation}

where $a$ ($a^{\dagger}$) and $b$ ($b^{\dagger}$) are the bosonic annihilation (creation) operators of the cavity mode (photon) and the mechanical degree of freedom (phonon), respectively; \( \Delta = \omega_c - \omega_L \) is the detuning between cavity ($\omega_c$) and laser ($\omega_L$) frequencies, \( \omega_m \) is the mechanical resonance frequency, and \( g_m > 0 \) denotes the OM coupling strength. In Fig.~\ref {fig:Figure2}(a), we show a representation of the OM system. In the Heisenberg picture, by adding dissipative effects, the corresponding equations of motion read

\begin{eqnarray}
    \dot{a} &=& -i\Delta a + ig_m (b + b^{\dagger}) + E(t) - \kappa a, \\
    \dot{b} &=& i\omega_m b + ig_m a^{\dagger}a - \gamma_m b,
\end{eqnarray}

where \( \kappa  \) and \( \gamma_m = \omega_m/Q \) are the cavity and mechanical damping rates, respectively, with \( Q\) being the mechanical quality factor. We initialize the OM system in the vacuum state, a common and experimentally realistic choice. This ensures that OM memory effects arise solely from the intrinsic nonlinear dynamics induced by the pulsed driving, rather than previous excitations. Under this assumption, the mean photon and phonon numbers read as

\begin{eqnarray}
   \langle a^{\dagger}a \rangle(t) &=& \int_{0}^{t} E(t')\, F_1(t,t',E)\, dt', \quad \textrm{(photon)}\label{n_cav}\\
   \langle b^{\dagger}b \rangle(t) &=& g_m \int_{0}^{t} F_2(t,t',E)\, dt', \quad\quad \textrm{(phonon)} \label{n_mec}
\end{eqnarray}

with the response kernels

\begin{eqnarray}
F_1(t,t',E) &=& \langle a + a^{\dagger} \rangle(t')\, e^{-2\kappa |t - t'|}, \\
F_2(t,t',E) &=& i \langle a^{\dagger}a(b^{\dagger} - b) \rangle(t')\, e^{-2\gamma_m |t - t'|}.
\end{eqnarray}

Note that $F_1$ depends only on photonic observables while $F_2$ is a hybrid function, where photonic and phononic observables play a role. The expectation values are evaluated in the Heisenberg picture, thus \( \langle A \rangle(t) \equiv \langle \Psi(0) | A(t) | \Psi(0) \rangle \) for operators $A(t)$ in this picture. The kernels \( F_{1,2}(t,t',E) \) depend implicitly on \( E(t') \) via the system's dynamical response.

\begin{figure*}[ht!]
\centering
\includegraphics[width = 0.85 \linewidth]{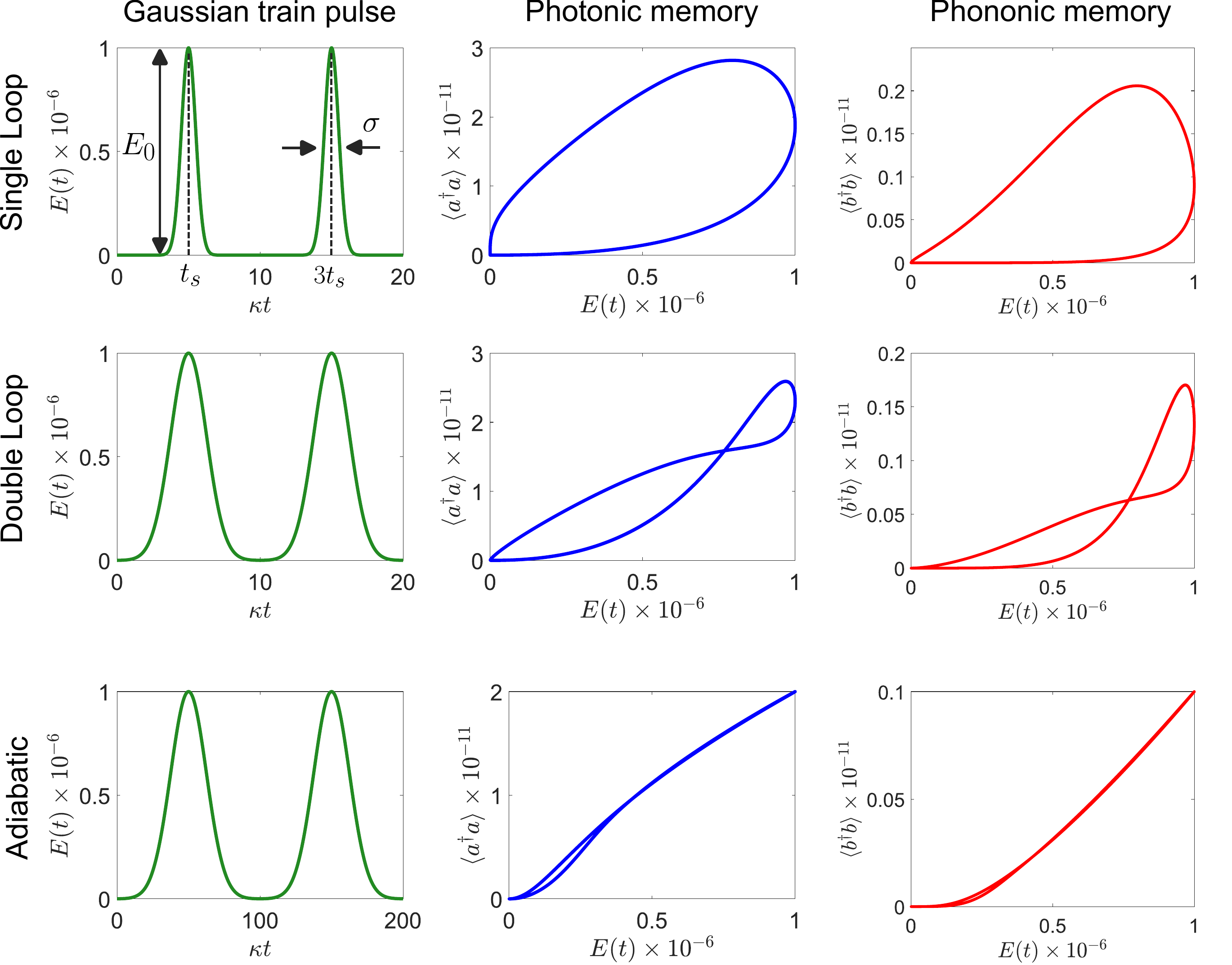}
\caption{Memory responses induced by the Gaussian train pulse defined in Eq.~\eqref{GaussianTrain} using $E_0 = 10^{4} \kappa$, $\omega_m = 20 \kappa$ and $Q=10^4$. From left to right: Gaussian train pulse, mean number of photons $\langle a^{\dagger}a\rangle$ and mean number of phonons $\langle b^{\dagger}b\rangle$ as a function of the input $E(t)$, respectively. From top to bottom: narrow and fast control ($\sigma = t_s/10$, $t_s = 5/\kappa$), broader and fast control ($\sigma = t_s/4$, $t_s = 5/\kappa$), and adiabatic response ($t_s = 50/\kappa$, $\sigma = t_s/4$). Here, memory effects are not accompanied by energy storage due to the choice of the control functions and parameters.}
\label{fig:Figure3}
\end{figure*}

Eqs.~\eqref{n_cav} and \eqref{n_mec} represent a time non-local (memory) response, where the observables \( y(t) \in \{\langle a^{\dagger}a \rangle, \langle b^{\dagger}b \rangle \} \) are determined by the control field \( E(t) \). In the decoupled limit \( g_m = 0 \), the optical and mechanical modes evolve independently, and the system reduces to a driven photonic memory element consistent with the Di Ventra-Pershin framework~\cite{DiVentraBook2023}. Here, dissipation introduces memory via convolution-like kernels with exponential decay terms $e^{-\Gamma|t-t'|}$ ($\Gamma =\{\kappa,\gamma_m \}$), reflecting the system’s intrinsic response timescales.

In typical OM experimental realizations, the decay rates \(\kappa\) and \(\gamma_m\) differ by several orders of magnitude. For instance, \(\kappa/2\pi\) usually ranges from \(10^5\) to \(10^7\) Hz, corresponding to optical quality factors \(Q_{\rm opt} \sim (10^5{-}10^7)\), while \(\gamma_m/2\pi\) spans between a few Hz to several kHz, with mechanical quality factors \(Q \sim (10^4{-}10^7)\)~\cite{Aspelmeyer2014,Kippenberg2008}. We set $Q = 10^6$ in this work to realize realistic numerical simulations. 

Although one might expect the mechanical mode to decay much more slowly than the optical mode, the optomechanical coupling alters this intuition. Much like two coupled oscillators with unequal damping rates, where the more highly damped oscillator accelerates the dissipation of the less damped one, here the rapid photonic dissipation (set by \(\kappa\)) drives the joint relaxation dynamics of the coupled system. As a result, the overall memory dynamics are governed by the cavity field timescale, and the most pronounced memory effects emerge at short times before photonic excitations have decayed.

\section{Mean-Field Approximation}

We focus on the large photon number regime, which is the state-of-the-art in OM setups. Experimental implementations in such directions include Fabry-P\'erot cavities with movable mirrors~\cite{Aspelmeyer2014} and microtoroidal resonators coupled to optical fibers~\cite{Vahala2003,Vahala2008}. In such systems, the OM coupling strength is small, \( g_m \sim( 1-10)\,\mathrm{GHz} \)~\cite{Aspelmeyer2014}, which requires working in the regime \( \langle a^{\dagger}a \rangle \gg 1 \) to observe significant effects. Under strong, coherent driving, it is well established that a mean-field approximation accurately describes the OM dynamics~\cite{OM_BHe1}. We define the dimensionless quadrature operators for the cavity and mechanical modes as
\begin{eqnarray}
 x_c &=& \frac{\left(a^{\dagger} + a\right)}{\sqrt{2}}, \quad p_c = i\frac{\left(a^{\dagger} - a\right)}{\sqrt{2}}, \\
 x_m &=& \frac{\left(b^{\dagger} + b\right)}{\sqrt{2}}, \quad p_m = i\frac{\left(b^{\dagger} - b\right)}{\sqrt{2}},
\end{eqnarray}
where $c$ and $m$ are the labels for cavity and mechanical modes, respectively. Here, we have the usual bosonic commutation relations \( [a,a^\dagger] = [b,b^\dagger] = 1 \). Their expectation values are denoted by capital letters: \( X_c = \langle x_c \rangle \), \( P_c = \langle p_c \rangle \), \( X_m = \langle x_m \rangle \), and \( P_m = \langle p_m \rangle \). In the mean-field approximation, operators are represented by $A = \langle A \rangle+\delta A$, where quantum fluctuations $\delta A$ are neglected, leading to \( \langle AB \rangle \approx \langle A \rangle \langle B \rangle \). The following set of nonlinear differential equations governs the resulting mean-field dynamics
\begin{eqnarray}
\dot{X}_c &=& -\kappa X_c + \Delta P_c - \sqrt{2} g_m X_m P_c + \sqrt{2} E(t), \label{OM1} \\
\dot{P}_c &=& -\kappa P_c - \Delta X_c + \sqrt{2} g_m X_m X_c, \label{OM2} \\
\dot{X}_m &=& -\gamma_m X_m + \omega_m P_m, \label{OM3} \\
\dot{P}_m &=& -\gamma_m P_m - \omega_m X_m + \frac{g_m}{\sqrt{2}} \left( X_c^2 + P_c^2 \right), \label{OM4}
\end{eqnarray}
where \( \langle a^{\dagger} a \rangle = (X_c^2 + P_c^2)/2 \) and \( \langle b^{\dagger} b \rangle = (X_m^2 + P_m^2)/2 \). Note that these equations are written in the rotating frame of the laser frequency \( \omega_L \), such that the detuning \( \Delta = \omega_c - \omega_L \) appears in the equations of motion but not in the drive \( E(t) \), differing from previous approaches~\cite{OM_BHe1,OM_Bhe8_BS}. In connection with Section~\ref{MemoryEffects}, the state vector is defined as $\mathbf{s} = (X_c, P_c, X_m, P_m) \in \mathbb{R}^4$ and the evolution function $g(\cdot)$ is determined by set of Eqs.~\eqref{OM1}-\eqref{OM4}.

We consider the resonant driving condition $\Delta = 0$ as it maximizes the OM memory effects under typical experimental conditions. The observables of interest can be expressed as time non-local responses
\begin{eqnarray}
   \langle a^{\dagger}a \rangle(t) &=& \sqrt{2} \int_{0}^{t} E(t') X_c(t') e^{-2\kappa|t - t'|} dt', \label{n_cav-MF} \\
   \langle b^{\dagger}b \rangle(t) &=& \sqrt{2} g_m \int_{0}^{t} \langle a^{\dagger}a \rangle(t') P_m(t') e^{-2\gamma_m |t - t'|} dt', \label{n_mec-MF}
\end{eqnarray}

where the factor $\sqrt{2}$ appears because of the definition of the dimensionless quadratures. In our simulations, we initialize the OM system in vacuum \( X_c(0) = P_c(0) = X_m(0) = P_m(0) = 0 \). Since the equations have no closed-form solution, we solve them numerically using the Runge-Kutta-Fehlberg method (ode45 in MATLAB), suitable for stiff, nonlinear dynamics.

\section{Control Engineering and Energy-Storing Response} \label{Control-Energy-Storing}

In this work, we explore memory responses in OM systems under control protocols commonly employed in quantum technologies. We consider three representative driving fields
\begin{eqnarray}
    E(t) &=& \sum_{n\, \text{odd}} E_0 e^{-(t-t_n)^2/(2\sigma^2)}, \; \text{(Gaussian train)}, \label{GaussianTrain}\\
    E(t) &=& E_0 \sin(\omega t), \hspace{1.8 cm} \text{(sinusoidal)}, \label{Sin}\\
    E(t) &=& E_0 \sin^2(\omega t), \hspace{1.6 cm} \text{(square-sinusoidal)}. \label{Square-Sin}
\end{eqnarray}
The Gaussian train pulse is parameterized by the amplitude \( E_0 \), pulse separation \( t_n = n t_s \), and width \( \sigma \), with periodicity \( T = 2t_s \). This control enables precise modulation of energy delivery and memory onset, as discussed in Sec.~\ref{MemoryGaussianPulse}. The sinusoidal drive, inspired by Floquet control techniques, depends on the amplitude \( E_0 \) and frequency \( \omega \), with period \( T = 2\pi/\omega \). Its sign-alternating nature leads to pinched-like hysteresis loops, characteristic of memristive responses. In contrast, the square-sinusoidal drive, with the same parameter set, yields a positive-definite signal of period \( T = \pi/\omega \), introducing sharp inflection points that enhance temporal asymmetry and energy injection.

A remarkable feature of this OM memory platform is its capacity to exhibit both energy-storing and energy non-storing behaviors, governed solely by the temporal structure of the driving field. This distinction is illustrated in Fig.~\ref{fig:Figure2}(b)-(c). Under a train of narrow and non-adiabatic Gaussian pulses (\( E_0 = 10^6 \kappa \), \( t_s = 5/\kappa \), \( \sigma = t_s/10 \)), the system evolves through closed input-output loops that intersect the origin, as shown in Fig.~\ref{fig:Figure2}(b). In this regime, the output observables, such as the mean photon number $\langle a^{\dagger}a\rangle= (X_c^2+P_c^2)/2$, vanish whenever the input does, indicating that all injected energy is dissipated within a single cycle. This defines an energy non-storing response, where no residual excitation remains to influence future dynamics.

In contrast, when driven by a square-sinusoidal field (\( E_0 = 10^6 \kappa \), \( \omega = 0.075 \omega_m \)), the system exhibits an energy-storing behavior, as shown in Fig.~\ref{fig:Figure2}(c), the output remains finite even when the input returns to zero, and the input-output trajectory encloses a finite area that does not pass through the origin. One expects synchronized input-output dynamics in the non-storing regime; when \( x(t) \to 0 \), the natural response is \( y(t) \to 0 \), as confirmed in Fig.~\ref{fig:Figure2}(d), where the photonic quadratures ($X_c, P_c$) decay to zero after each driving cycle.

Strikingly, under square-sinusoidal driving, the system may enter a regime where input and output break the previous condition. For a specific range of frequencies, the photon momentum \( P_c \) is close to its maximum when \( E(t) \) is near its minimum, and remains nonzero after the field vanishes, as shown in Fig.~\ref{fig:Figure2}(e), where the contribution of $X_c$ is less dominant. This phase mismatch between quadratures and input is a key signature of the energy-storing regime, highlighting the essential role of temporal asymmetry and internal dynamical inertia in generating persistent memory effects in cavity OM systems.

\section{Memory Effects via Gaussian Train Pulse Engineering}\label{MemoryGaussianPulse}

To explore the diversity of memory responses in the OM system, we now consider the Gaussian train pulse defined in Eq.~\eqref{GaussianTrain}. This control reduces to a sequence of $\delta$-like kicks in the limit \( \sigma \to 0 \), corresponding to the strongly non-adiabatic regime, discussed in Appendix~\ref{Delta-Pulse}. Conversely, for sufficiently slow driving (\( t_s \gtrsim 50/\kappa \)), the system evolves quasi-adiabatically, mimicking memoryless behavior.

\begin{figure}[ht!]
\centering
\includegraphics[width = 1 \linewidth]{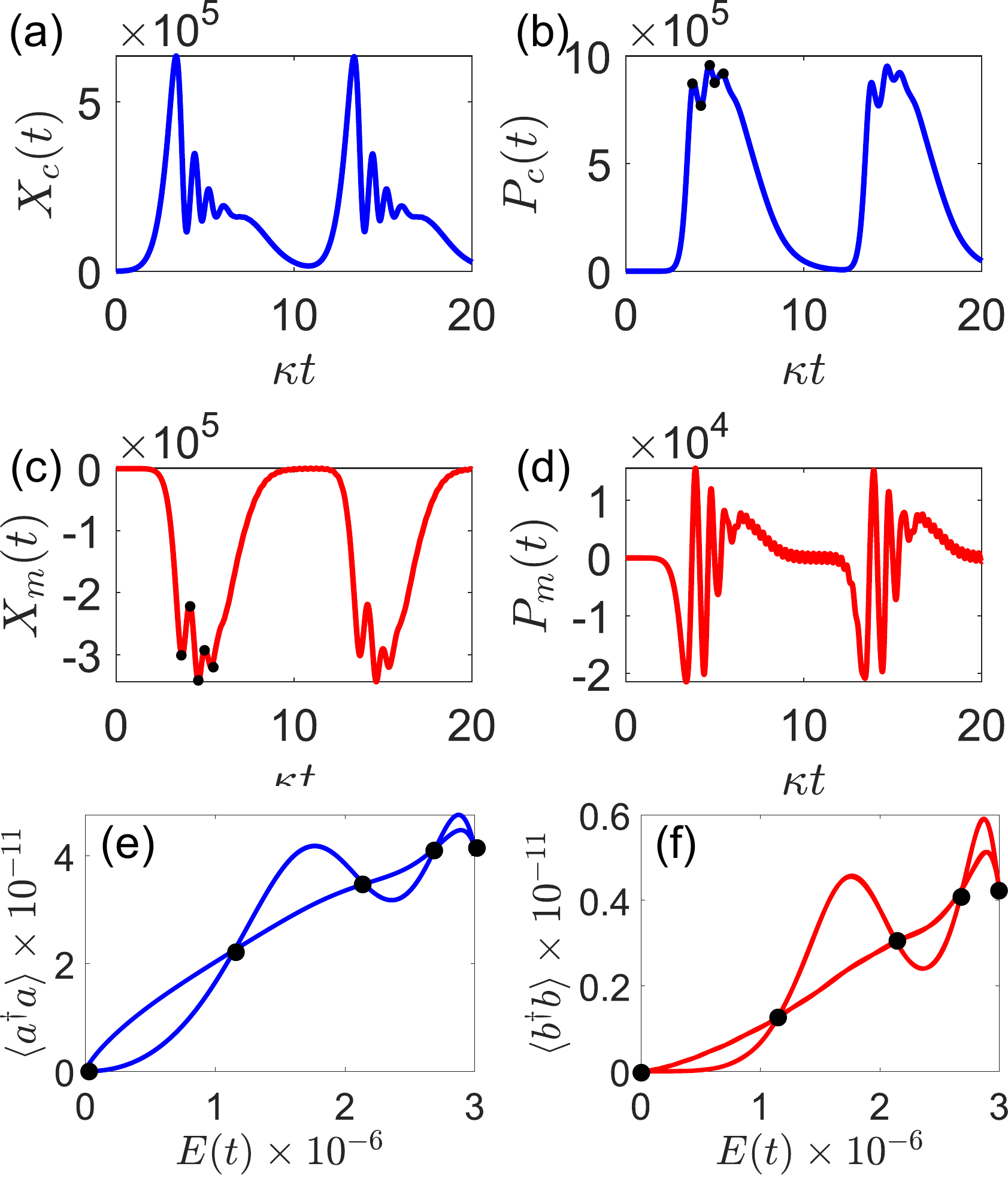}
\caption{\texorpdfstring{$n$}{n}-loop hysteresis induced by the Gaussian train pulse ($n= 5$). (a) and (b) are the photonic dimensionless quadratures, while (c) and (d) are the phononic dimensionless quadratures. Memory effects are observed in the dynamical hysteresis curves for the mean number of (e) photons and (f) phonons. For this simulation, we use a train pulse of Gaussian functions with $\omega_m = 20\kappa$, $Q=10^4$, $g_m = 10^{-5}\kappa$, $E_0 = 3\times 10^6\kappa$, $\sigma = t_s/4$, and $t_s = 5\kappa$.}
\label{fig:Figure4}
\end{figure}

The Gaussian amplitude \( E_0 \) plays a subtler role in this nonlinear setting. Under strong and continuous driving, optomechanical systems can exhibit amplitude locking, where the mean phonon number stabilizes at discrete values~\cite{OM_BHe1,OM_BheMO_frozen}. In this regime, the mechanical displacement approximately follows \( X_m(t) \approx A \cos(\omega_m t + \phi) + d \), with \( d \ll A \), leading to quantized plateaus in \( \langle b^\dagger b \rangle \) as a function of \( E_0/\kappa \), typically in the range \( E_0/\kappa \approx (0.2 - 3.5)\times 10^7 \). Recently, anomalous stabilization mechanisms were reported in coupled cavity-mechanical dynamics~\cite{BHe2024}. These phenomena indicate a broad range of nonlinear memory behaviors that can be accessed by tuning the parameters \( (E_0, \sigma, t_s) \) properly.

\begin{figure*}[ht!]
\centering
\includegraphics[width = 1 \linewidth]{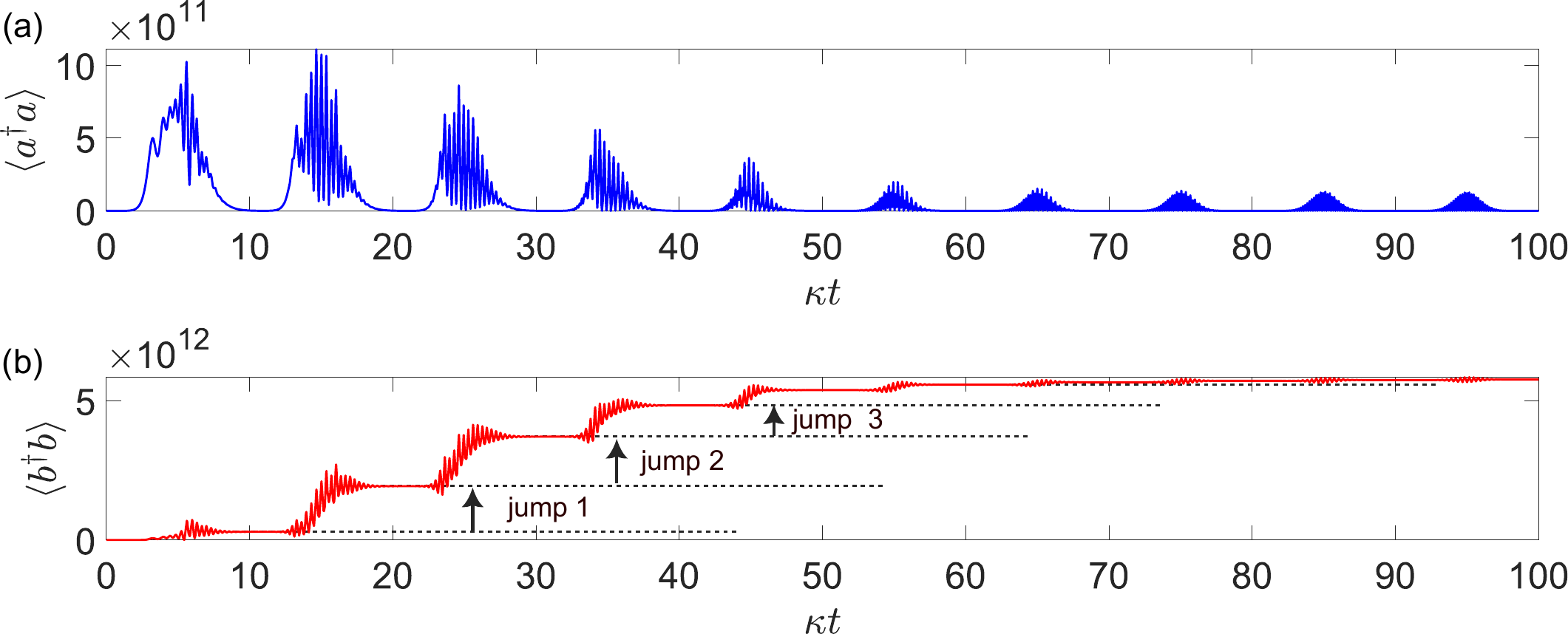}
\caption{Dynamical jumps in the mean number of phonons induced by a Gaussian train pulse in the large amplitude regime. Temporal evolution of (a) photon and (b) phonon populations under high-amplitude pulsing. Parameters: $\omega_m = 20\kappa$, $Q=10^4$, $g_m = 10^{-5}\kappa$, $E_0 = 6\times 10^6\kappa$, $\sigma = t_s/4$, and $t_s = 5/\kappa$.}
\label{fig:Figure5}
\end{figure*}

In Fig.~\ref{fig:Figure3}, we present three representative dynamical responses using input-output trajectories \( \gamma(t) = (x(t), y(t)) \), with \( x(t) = E(t) \) and \( y(t) \in \{\langle a^\dagger a \rangle, \langle b^\dagger b \rangle\} \). Each panel shows different Gaussian control regimes and the corresponding photonic/phononic observables, revealing hysteresis loops under fast and narrow pulses and reversible dynamics in the adiabatic limit. In what follows, we separate the analysis of the memory responses observed in this device under the Gaussian effect, corresponding to the energy non-storing case.

\subsection{Single-loop hysteresis}

As shown in the upper panel of Fig.~\ref{fig:Figure3}, the single-loop regime features a smooth, closed input-output trajectory in the first quadrant, free of self-intersections. This behavior arises under moderate amplitudes (\( E_0 \lesssim 10^6 \kappa \)), narrow Gaussian pulses (\( \sigma \lesssim t_s/5 \)), and relatively fast control times (\( t_s \sim (1-10)\,\kappa^{-1} \)). The enclosed area is finite and signals memory, similar to the behavior of classical memristive elements. This demonstrates that the optomechanical system can emulate tunable memristive responses through pulse shaping.

\subsection{Double-loop hysteresis}

As the pulse width \( \sigma \) increases while keeping \( t_s \) fixed (fast control), the system transitions into a double-loop regime, illustrated in the middle panel of Fig.~\ref{fig:Figure3}. The input-output curve develops a self-intersection, forming two lobes. This behavior arises in an intermediate regime, where the control is fast but sufficiently broad to excite internal oscillations. As shown in Fig.~\ref{fig:Figure4}, oscillatory modulations in the quadratures lead to nodes in the loop. Such features indicate more complex memory responses, possibly linked to bistable or multi-state dynamics.

\subsection{Adiabatic regime}

In the bottom panel of Fig.~\ref{fig:Figure3}, the adiabatic regime is reached for slow, wide pulses (\( t_s \gtrsim 50/\kappa \)). Here, the system evolves quasi-statically, and the input-output loop collapses into a narrow curve with negligible area, consistent with a memoryless response. In this limit, the system behaves as an Ohmic material \( y(t) = R(t)x(t) \), with no hysteresis or history-dependent effects. This highlights the key role of fast, non-adiabatic modulation in inducing memory.

\subsection{\texorpdfstring{$n$}{n}-loop hysteresis}

Increasing the amplitude to \( E_0 \sim (1{-}4) \times 10^6 \kappa \) while maintaining fast driving (\( t_s < 10/\kappa \)) leads to multi-loop hysteresis, as seen in Fig.~\ref{fig:Figure4}. These curves exhibit \( n > 2 \) self-intersections, with \( n \) reflecting the number of oscillations in the photonic and phononic quadratures. The number of nodes correlates with peaks in the quadrature \( P_c(t) \) for photons and \( X_m(t) \) for phonons, indicating intricate memory states distributed across multiple cycles. This regime showcases the system’s capacity for complex, nonlinear input-output behavior.

\subsection{Dynamical quantized jumps in the mechanical mode}

The system exhibits sudden transitions in the mean phonon number when the amplitude increases beyond \( E_0 \gtrsim 5 \times 10^6\kappa \). For instance, with \( E_0 = 6 \times 10^6 \kappa \), \( t_s = 5/\kappa \), and \( \sigma = t_s/4 \), we observe dynamical discrete jumps in \( \langle b^\dagger b \rangle \), as shown in Fig.~\ref{fig:Figure5}. These transitions are accompanied by decaying oscillatory packets in $\langle a^{\dagger}a\rangle$, suggesting a regime where chaotic dynamics coexist with quantized plateaus. Such behavior echoes amplitude-locking phenomena reported in Refs.~\cite{OM_BHe1, OM_BheMO_frozen, BHe2024}, highlighting the rich interplay between nonlinearity, dissipation, and pulsed excitation in optomechanical platforms. These findings open new directions for exploring multistable and stochastic memory effects. 

In the following section, we introduce two additional driving, sinusoidal and square-sinusoidal pulses, and analyze their performance using the form factor from Eq.~\eqref{FormFactor}, to benchmark the memory capacity of different control schemes against several quantum memristive systems.

\section{Optimal memory response under periodic driving}

We now investigate the optimal memory performance of the optomechanical setup under the periodic driving protocols \( E(t) \) given in Eqs.~\eqref{GaussianTrain}-\eqref{Square-Sin}. The figure of merit is the dimensionless form factor \(\mathcal{F}\), which captures the geometric memory content of the input-output trajectory, as defined in Eq.~\eqref{FormFactor}. Optimizing \(\mathcal{F}\) has been previously studied in the context of single and coupled quantum memristors based on circuit QED architectures~\cite{Hernani2024}. This optimization is relevant for memory-based technologies such as neuromorphic computing and quantum memdevices, where the efficient encoding of time-correlated information is crucial.

Here, we explore how memory can be maximized in OM systems by tuning experimentally accessible parameters of the driving fields. We compare three periodic controls (Gaussian trains, sinusoidal, and square-sinusoidal pulses) and report the optimal form factor achieved for each case. In the following subsections, we describe the optimization procedure and analyze the results obtained for different driving functions and other reported works.

\subsection{Form factor optimization}

The optimization protocol proceeds as follows. First, we define a periodic driving field \( E(t) \) satisfying \( E(t + T) = E(t) \), where \( T \) is the period. The control function is parametrized by a set of real-valued variables \( \boldsymbol{\theta} \in \mathbb{R}^d \), with \( d \) denoting the number of tunable parameters. For instance, the sinusoidal and square-sinusoidal drives on Eq.~\eqref{Sin} have \( \boldsymbol{\theta}  = (E_0, \omega_d) \) and \( d = 2 \). The Gaussian train pulse given in Eq.\eqref{GaussianTrain}, implies \( \boldsymbol{\theta}  = (E_0, t_s, \sigma) \), with \( d = 3 \).

\begin{figure*}[ht!]
\centering
\includegraphics[width = 0.85 \linewidth]{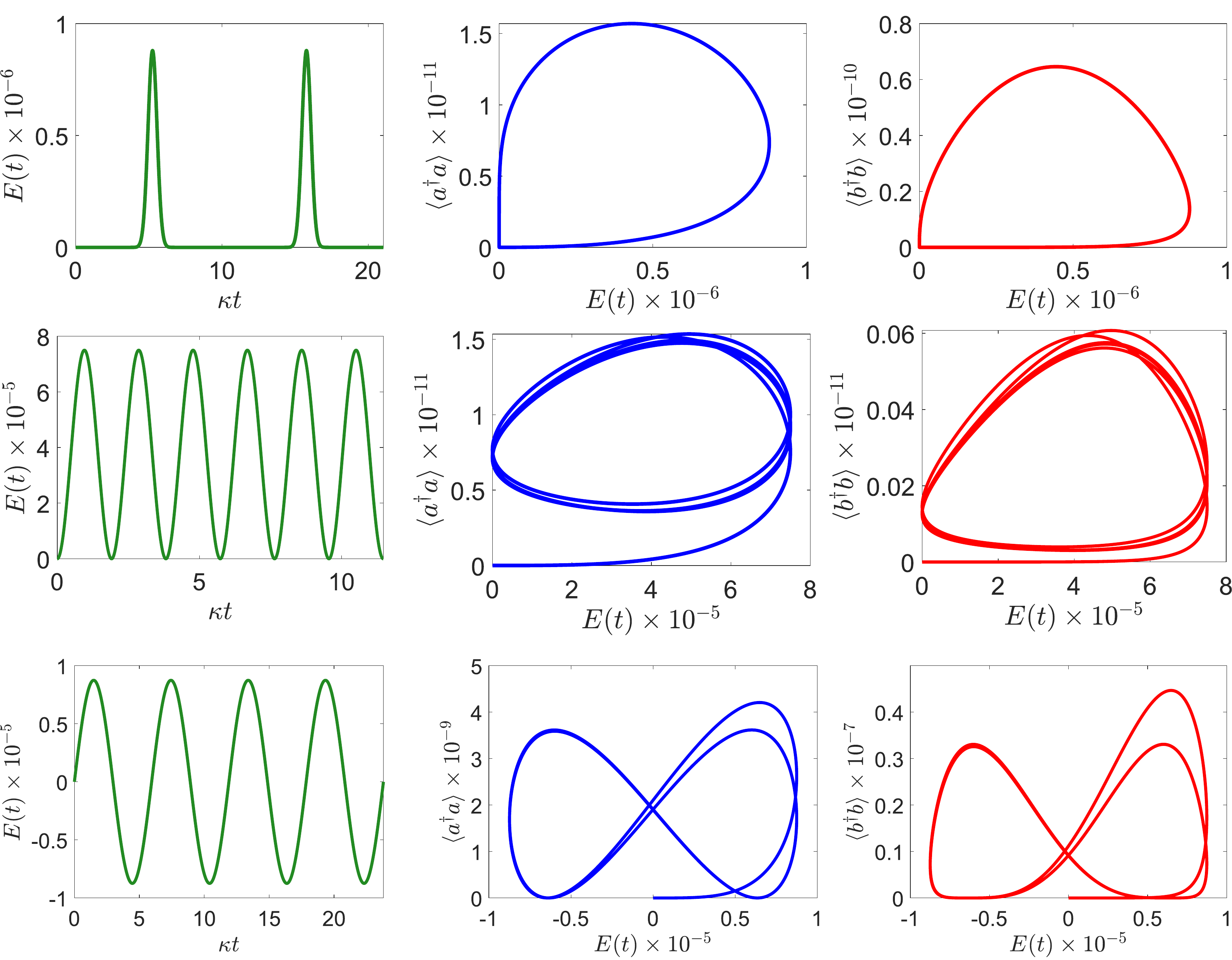}
\caption{Optimal memory responses via maximization of the form factor \(\mathcal{F}\) using the mean number of photons as output and calculated for different periodic pulses $E(t)$: (from top to bottom) Gaussian train, square-sinusoidal, and sinusoidal control fields. The control parameters were optimized using the genetic algorithm. The corresponding input-output trajectories are plotted in normalized units \((\widetilde{x}(t), \widetilde{y}(t))\) to facilitate comparison.}
\label{fig:Figure6}
\end{figure*}

Second, we numerically solve the nonlinear mean-field equations~\eqref{OM1}-\eqref{OM4} using an initial trial configuration \( \boldsymbol{\theta} _0 \). The objective is to find the optimal configuration \( \boldsymbol{\theta} ^\star \) that maximizes the average form factor over \( N \) driving cycles. Specifically, we evaluate the form factor \(\mathcal{F}(\boldsymbol{\theta}, t_n)\) at the end of each cycle \( t_n = nT \), with \( n = 1, 2, \dots, N \), and define the cost function
\begin{equation} \label{CostFunction}
    C(\boldsymbol{\theta} ) = -\frac{1}{N} \sum_{n=1}^N \mathcal{F}(\boldsymbol{\theta}, t_n),
\end{equation}
so that minimizing \( C(\boldsymbol{\theta}) \) corresponds to maximizing memory content. The form factor at cycle \( n \) is computed using
\begin{eqnarray}
    \mathcal{F}(\boldsymbol{\theta}, t_n) &=& \frac{4 \pi A(\boldsymbol{\theta}, t_n)}{P^2(\boldsymbol{\theta}, t_n)}, \\
    A(\boldsymbol{\theta}, t_n) &=& \left| \int_{t_{n-1}}^{t_n} E(\boldsymbol{\theta},t)\, \dot{y}(\boldsymbol{\theta}, t)\, dt \right|, \\
    P(\boldsymbol{\theta}, t_n) &=& \int_{t_{n-1}}^{t_n} \sqrt{ \dot{E}^2(\boldsymbol{\theta},t) + \dot{y}^2(\boldsymbol{\theta}, t)} \, dt,
\end{eqnarray}
where \( y(\boldsymbol{\theta}, t) \) is the observable used to define the input-output trajectory, and \( t_0 = 0 \) is the initial time. All integrals are evaluated numerically using the trapezoidal rule.

To estimate the global minimum of the cost function \( C(\boldsymbol{\theta}) \), we use a genetic algorithm with a population size of \( 50 + 10 \times d \), a Gaussian mutation scheme, and a maximum of 100 generations. The mathematical foundation of the genetic algorithm is presented in Ref.~\cite{Conn1997}. Genetic algorithms are particularly well-suited for exploring high-dimensional, nonlinear landscapes. As discussed in Ref.~\cite{Hernani2024}, alternative approaches based on machine learning can also be employed for this task. In our case, we consider the input \( x(t) = E(t) \) and explore different observables for the output \( y(t) \), including the mean photon and phonon numbers.

\subsection{Comparative analysis of optimized periodic drives}

In Fig.~\ref{fig:Figure6}, we present the photonic optimized hysteresis curves obtained by applying our optimization protocol to the cost function~\eqref{CostFunction}. From top to bottom, we display the optimal solutions for the Gaussian train, sinusoidal, and square-sinusoidal pulses, respectively. The results, summarized in Table~\ref{tab:formfactor}, reveal differences in memory performance among the three classes of periodic drivings, as quantified by the optimal form factor $\mathcal{F}_{\rm opt}$, and also by making a comparison with quantum platforms analyzed in the literature. The same optimization protocol can be applied by using the mean number of phonons as the output; such results are reported in Table~\ref{tab:formfactor}. 

In the OM system, the square-sinusoidal protocol yields the highest value for single-loop hysteresis for both photons and phonons, reaching \(\mathcal{F}_{\rm opt} = 0.965\) and \(\mathcal{F}_{\rm opt} = 0.963\), respectively. These optimal values are closely followed by the Gaussian train with \(\mathcal{F}_{\rm opt} = 0.925\) for photons and a smaller value for phonons \(\mathcal{F}_{\rm opt} = 0.863\). Memristive curves are experimentally observed in photonic quantum memristors using a tunable beamsplitter~\cite{PQM_1experimental}, where the mean number of photons quantifies memory effects. By optimizing the form factor of Ref.~\cite{PQM_1experimental}, we obtain the optimal form factor close to $0.58$. The latter observation suggests that our protocol outperforms the optimal form factor of previous photonic setups.

Hybrid quantum systems, such as polariton-based quantum memristors, offer a unique plasticity property,  the input-output loop depends on the quantum state
initialization~\cite{Qmem_Polariton}. However, according to our numerical simulations, optimizing the form factor in this polaritonic device leads to optimal form factors around $0.2$. Therefore, compared with current hybrid platforms, our OM device exhibits better results regarding the geometrical characterization of memory. 

\begin{table*}[ht!]
\centering
\caption{
Optimized form factor for different periodic driving protocols in our optomechanical model and comparison with selected platforms from the literature. Input-output trajectories were normalized as $\widetilde{x}(t) = x(t)/\max[x(t)]$ and $\widetilde{y}(t) = y(t)/\max[y(t)]$. Simulations use $\omega_m = 20\kappa$, $\gamma_m = \omega_m/Q$, $Q = 10^4$, and $\Delta = 0$. Optimal parameters marked with the symbol ``$^{\ast}$'' denote that the optimization was realized in this work. Theoretical and experimental realizations are denoted in parentheses using (Theo.) and (Exp.), respectively.}
\begin{tabular}{p{4.5 cm} p{4.0 cm} p{6.0 cm} c}
\hline\hline
\textbf{Platform} & \textbf{Mechanism} & \textbf{Control parameters} & \textbf{Form factor}/\textbf{Max} \\
\hline
\multirow{6}{=}{Cavity optomechanics (Theo.)} 
& \multirow{6}{=}{Nonlinear and pulsed optomechanical dynamics in the large photon and dissipative regime} 
& \textbf{Gaussian (photons)}: $E_0 = 5.717\times10^5\,\kappa$, $t_s = 16.119\,\kappa^{-1}$, $\sigma = 0.313\,\kappa^{-1}$ 
& $0.925^{\ast}/1$ \\
& & \textbf{Gaussian (phonons)}: $E_0 = 2.015\times10^5\,\kappa$, $t_s = 30.974\,\kappa^{-1}$, $\sigma = 0.224\,\kappa^{-1}$ 
& $0.863^{\ast}/1$\\
\cline{3-4}
& & \textbf{Sinusoidal (photons)}: $E_0 = 8.745\times10^4\,\kappa$, $\omega = 1.055\,\kappa$ 
& $0.450^{\ast}/0.5$ \\
& & \textbf{Sinusoidal (phonons)}: $E_0 = 7.895\times10^4\,\kappa$, $\omega = 1.918\,\kappa$ 
& $0.441^{\ast}/0.5$ \\
\cline{3-4}
& & \textbf{Square-sinusoidal (photons)}: $E_0 = 7.498\times10^5\,\kappa$, $\omega = 1.644\,\kappa$ 
& $0.965^{\ast}/1$ \\
& & \textbf{Square-sinusoidal (phonons)}: $E_0 = 2.173\times10^5\,\kappa$, $\omega = 2.794\,\kappa$ 
& $0.963^{\ast}/1$ \\
\hline
Circuit QED~\cite{Hernani2024} (Theo.)
& Non-Markovian open dynamics and entanglement oscillation under magnetic flux control 
& Control: external magnetic field flux $\phi_{dl}(t) = \phi_0 + A \sin(\omega_l t)$. Input: quasi-particle current $I = G_l(t)\hat{V}_{\text{cap},l}(t)$. Output: Voltage across the capacitor: $\hat{V}_{\text{cap},l}(t)=-2e\langle \hat{n}_l(t) \rangle/C_{\Sigma,l}$
& 
\(
\begin{array}{@{}l@{}}
0.324/0.5 \; \text{(single)} \\
0.503/0.5 \; \text{(double)}
\end{array}
\) \\
Quantum polaritons~\cite{Qmem_Polariton} (Theo.)
& Atomic modulation with polaritonic delocalization in the Jaynes-Cummings-Hubbard model
& Input/Control: atomic modulation $\xi(t) = \xi_i + (\xi_f-\xi_i)e^{-(t-T)^2/(2\sigma_w^2)}$. Output: variance in the polariton number $\mbox{var}[N_i] = \mbox{Tr}[N_i^2 \rho]-\mbox{Tr}[N_i \rho]^2$ with $N_i = a_i^{\dagger}a_i + \sigma_i^{\dagger}\sigma_i$
& $0.192^{\ast}/1$ \\
Quantum photonic memristor~\cite{PQM_1experimental} (Exp.)
& Tunable reflectivity via integrated photonics
& Input/Control: $\langle n_{\rm in}(t)\rangle = \sin^2(\pi/T_{\rm osc} t)$ with optimal $T \approx 0.37 T_{\rm osc}$. Output: $\langle n_{\rm out}(t)\rangle = [1-R(t)]\langle n_{\rm in}(t)\rangle$, with tunable reflectivity $R(t)$
& $0.58^{\ast}/1$ \\
Quantum photonic memristor~\cite{Ferrara2025}  (Theo.)
& Entangled two-photons using tunable reflectivity in a beamsplitter via integrated photonics 
& Input/Control: $\langle n_{\rm in}(t)\rangle = \sin^2(\pi/T_{\rm osc} t)$ with optimal $T \approx 0.37 T_{\rm osc}$. Output: concurrence and quantum coherence hysteresis cycles with tunable reflectivity $R(t)$
& \(
\begin{array}{@{}l@{}}
\approx 0.66/1 \; \text{(concurrence)} \\
\approx 0.62/1 \; \text{(coherence)}
\end{array}
\) \\
\hline\hline
\end{tabular}
\label{tab:formfactor}
\end{table*}

It is worth noting that sinusoidal drivings generally lead to pinched hysteresis curves, which satisfy $\mathcal{F} \leq 0.5$ by construction. Then, our sinusoidal drive achieves \(\mathcal{F}_{\rm opt} = 0.450\) and \(\mathcal{F}_{\rm opt} = 0.4441\) for photons and phonons, respectively, where the maximum form factor is $0.5$ for a pinched-like loop. As a comparison, in circuit QED memristors with pinched hysteresis~\cite{Hernani2024}, the optimal value for the average form factor is $0.32$ (single) and $0.5$ (coupled) for a superconducting device controlled by a sinusoidal magnetic flux. Therefore, the performance of our OM device is between single and double superconducting performance. 

These differences between our control schemes originate from the degree of temporal asymmetry and sharpness in the control profiles. Moreover, the optimal Gaussian train pulse leads to an energy non-storing response, contrary to optimal sinusoidal and square-sinusoidal control fields. The sharp temporal edges in the square-sinusoidal pulses induce strong nonlinearities and non-adiabatic responses, effectively enhancing the system's intrinsic memory capability. Similarly, Gaussian pulse trains behave as localized bursts of energy, pushing the system far from adiabaticity and preventing an instantaneous response. This delay leads to effective hysteresis and memory encoding, similar to a memristor. In contrast, even having a smoother derivative, the purely sinusoidal case produces high values for the optimal form factor.

A further distinction appears in the energy scales involved in the optimal form factor. The maximum photon number \(\langle a^{\dagger}a \rangle_{\rm max}\) varies by two orders of magnitude across protocols: sinusoidal driving produces values around \(\sim 4 \times 10^9\), while both Gaussian train and square-sinusoidal pulses exceed \(\sim (0.6-1.5)\times 10^{11}\). The latter can be explained by looking at the optimal values of the amplitude $E_0$ reported in Table~\ref{tab:formfactor}, where $E_0 \sim 8\times 10^4 \kappa$ for the sinusoidal control, and $E_0 \sim (6-8)\times 10^5 \kappa$ for Gaussian and square-sinusoidal fields. This indicates a correlation between energy injection and memory efficiency since Gaussian and square-sinusoidal fields give the maximum optimal form factor. 

From an experimental perspective, the presented pulses are viable and accessible using current optical modulation techniques. Their amplitude, frequency, and duration parameter sets can be finely tuned in state-of-the-art cavity optomechanical setups. Based on the form factor, our geometric optimization framework provides a physically intuitive and broadly applicable method to benchmark dynamical memory across different quantum platforms. This includes trapped ions~\cite{QMem_TrappedIon}, superconducting circuits~\cite{Hernani2024}, and polaritonic systems~\cite{Qmem_Polariton}, where hysteresis and history-dependent observables are reported.

\section{Forced Damped Oscillator Analogy}
In systems where the output takes the form \( y(t) = f(s,x,t)\, x(t) \), such as memristive devices, memory effects have often been interpreted through analogies with electrical circuits~\cite{Pershin2011}. While such perspectives have proven useful, for instance, in polariton-based quantum memristors~\cite{Qmem_Polariton}, the presence of a mechanical degree of freedom in cavity OM platforms naturally invites a mechanical analogy.

From the mean-field equations~\eqref{OM1}-\eqref{OM4}, the mechanical displacement \( x(t) = x_0 \langle b + b^\dagger \rangle \) obeys a second-order differential equation that closely resembles the dynamics of a driven, damped harmonic oscillator
\begin{equation} \label{FDO}
    m \ddot{x} + 2m\gamma_m \dot{x} + m\omega_0^2 x = f(t),
\end{equation}
where \( x_0 = \sqrt{\hbar / (2m\omega_m)} \) is the mechanical zero-point amplitude, and the effective frequency is given by \( \omega_0 = \sqrt{\omega_m^2 + \gamma_m^2} \), which includes a damping-induced shift. The resulting force is of a photonic nature
\begin{equation}
f(t) = -\left\langle \frac{\partial H_{\mathrm{int}}}{\partial x_m} \right\rangle = g_m \langle a^\dagger a \rangle
\end{equation}
and originates from the OM interaction Hamiltonian \( H_{\mathrm{int}} = -g_m a^\dagger a (b + b^\dagger) \), revealing how cavity photons mediate the actuation of the mechanical degree of freedom. Eq.~\eqref{FDO} highlights that the OM coupling \( g_m \) is the essential element triggering mechanical motion. In the limit \( g_m = 0 \), and for initial conditions \( x(0) = \dot{x}(0) = 0 \), the mechanical displacement remains null for all times. Thus, cavity excitations are solely responsible for generating mechanical motion and, in turn, memory effects in the OM system.

This mechanical analogy provides physical intuition into the system’s hysteresis and memory behavior and connects the dynamics of quantum memdevices with classical models of driven dissipation. Such insights can guide the design of OM protocols tuned for resonance-enhanced memory and nonlinear response.



\section{Conclusions}

We have developed a theoretical and computational framework to investigate diverse memory effects in cavity optomechanical systems subjected to periodic pulsed laser driving. By introducing a geometric measure of memory based on the area and perimeter of input-output trajectories, we have provided a physically transparent and experimentally accessible criterion for detecting and quantifying memory.

Our results demonstrate that the emergence of memory in these systems is intrinsically linked to the temporal asymmetry and the non-adiabatic response induced by the driving field. Through the mean-field simulations, we have shown that tailored pulse sequences, particularly the Gaussian trains, sinusoidal, and square-sinusoidal drives, can generate pronounced hysteresis and optimize the memory content, as quantified by the form factor. Among the protocols tested, the square-sinusoidal driving achieves the highest performance, while the Gaussian trains offer similar advantages with additional tunability. In particular, Gaussian engineering leads to diverse memory effects such as $n$-loop hysteresis, adiabatic responses, and dynamical jumps in the phonon number.

These findings position cavity optomechanics as a promising and versatile platform for implementing programmable memory at the quantum-classical interface. The framework presented here can be extended to other driven-dissipative systems, including polaritonic lattices, trapped ions, and circuit QED architectures. Future works should address problems related to system scalability, noise effects, and quantum corrections beyond the mean-field approximation. More broadly, our work lays the foundation for designing and optimizing OM memory devices, leveraging dynamical hysteresis and an interplay between the energy storing and non-storing responses via nonlinear dynamics, dissipation, and periodic pulsed control.

\section{Acknowledgments}
 H.T-M acknowledges support from the Universidad Mayor through the Doctoral fellowship. B.H. acknowledges the financial support from the project Fondecyt Regular No. 1221250. A.N. acknowledges the financial support from the project Fondecyt Regular No. 1251130. M.D. acknowledges support from the National Science Foundation grant No. ECCS-2229880.

\appendix

\section{Photonic memory induced by a Dirac $\delta$ pulse} \label{Delta-Pulse}

To illustrate the emergence of memory in the strong non-adiabatic regime, we consider a Dirac $\delta$ pulse \( E(t) = E_0 \delta(t - t_s) \) as a toy model. Here, \( E_0 = \int_0^\infty E(t')\,dt' \) is a dimensionless amplitude, and \( t_s \) is the pulse time. The period of this drive is defined as \( T = 2t_s \). We define the output observable as the cavity energy \(\langle H_c \rangle(t) = \omega_c \langle a^{\dagger}a \rangle(t) \). From Eq.~\eqref{n_cav}, we get

\begin{equation}
   \langle H_c \rangle (t) = H_0 e^{-2\kappa (t - t_s)} \Theta(t - t_s),
\end{equation}

where \( H_0 = \sqrt{2} \omega_c E_0^2 \) is the maximum cavity energy, and \( \Theta(t - t_s) \) is the Heaviside step function. The input \( x(t) = E(t) \) is a classical control field, symmetric around \( t = t_s \), \textit{i.e.}, \( x(t_s - t) = x(t_s + t) \). In contrast, the output \( y(t) = \langle H_c \rangle(t) \) is a quantum observable that exhibits a discontinuity at \( t = t_s \), quantified by
\begin{equation}
\lim_{\epsilon \to 0} \left[\langle H_c \rangle(t_s + \epsilon) - \langle H_c \rangle(t_s - \epsilon)\right] = H_0.
\end{equation}

The resulting input-output trajectory \( \gamma(t) = (x(t), y(t)) \) begins at \( \gamma(0) = (0, 0) \), and asymptotically returns to \( \gamma(T) \approx (0, 0) \) if \( 2\kappa T \gg 1 \), closing the cycle. The enclosed area can be computed via Green's theorem

\begin{equation}
A = \left| \int_{0}^{T} E(t) \, {d\langle H_c\rangle \over dt}  \, dt\right| = 2\sqrt{2} \kappa \omega_c E_0^3 e^{-2\kappa t_s} \neq 0,
\end{equation}

which confirms the presence of memory effects according to the geometric criterion. This toy model highlights two key insights: (i) the explicit breaking of time-reversal symmetry between input and output is a necessary (though not sufficient) condition for memory, and (ii) dissipation plays a critical role by extending the output response beyond the duration of the input. The OM system retains memory over a characteristic decay time \( \tau = 1/(2\kappa) \), even without continued driving. In fact, $A=0$ if $\kappa = 0$, revealing the relevance of dissipation.

\bibliographystyle{unsrt}

\end{document}